# Ghanaian Consumers' Online Privacy Concerns: Causes and its Effects on E-Commerce Adoption

E.T. Tchao
Dept. of Computer Engineering
Kwame Nkrumah Univ. Science and Tech
Kumasi - Ghana

Christiana Aggor
Department of Computer Engineering
Kwame Nkrumah Univ. Science and Tech
Kumasi - Ghana

Kwasi Diawuo
Department of Electrical and Electronic Engineering
University of Energy and Natural Resources
Sunyani – Ghana

Seth Djane Kotey
Dept. of Computer Engineering
Kwame Nkrumah University of Science and Tech
Kumasi - Ghana

*Abstract*—**Online privacy has gradually become a concern for internet users over the years as a result of the interconnection of customers' devices with other devices supporting the internet technology. This research investigates and discusses the factors that influence the privacy concerns faced by online consumers of internet services and the possible outcomes of these privacy concerns on the African online market with Ghana being the primary focus. Results from this study indicated that only 10.1% of respondents felt that the internet was safe for purchase and payment transaction in Ghana. However, respondents were willing to shop online if e-Commerce was the only means of getting their products. Respondents also had a high sense of perceived vulnerability and their perceived vulnerability to unauthorized data collection and misuse of personal information could affect Ghanaian e-Commerce platform adoption. The perceived ability of users of e-Commerce platforms in Ghana to control data collection and its subsequent use by other third parties was also found to negatively impact customers' willingness to wholly transact and share their personal information online. The perceived vulnerability was found to be affected by the high levels of internet illiteracy whiles the perceived ability to control the collection of information and use was influenced by both the internet literacy level as well as the level of social awareness of the Ghanaian internet consumer.**

## I. INTRODUCTION

Internet, which is defined as the linkage of networks, is used by a wide variety of people. From its early concept, it was supposed to be used by only the military and the government but now it has now extended in terms of usage to billions of people from all walks of life [1]. Since the inception of the world-wide web, the number of people who have been using the internet has grown exponentially. The internet has further changed a number of ways we do things from how we communicate, to the advancement of research and how businesses are transacted [2]. Thus, the internet has turned the earth planet as we know it into a "global village" whereby information has been made readily available to everyone.

The internet is rapidly evolving and two aspects have made the evolution a reality, that is, the breakthrough of mobile technology and the introduction of social networking. The internet never ceases to be the pinnacle of a new beginning and will undoubtedly continue to play an unprecedented role in our life. A report from authors in [3] points out that the number of users of the internet were roughly 16 million in 1995 and in 2017 the number is now 3.732 billion which represents 49.6% of the entire world's population.

The ever increasing number of users of the internet technology however creates one major concern and that is privacy. UNESCO defines privacy as the right of any citizen to control his or her own personal information and to decide what that information can be used for, thus, to keep or disclose [4]. Privacy invasion comes into the picture when the personal information is being disseminated, used and released without being restrained. With the rapid expansion of the world-wide web, consumers have raised concerns regarding their personal privacy whilst they are either online or even offline.

The internet in spite of its numerous advantages has paved way for a cleverly systemized way of collecting data. The online footprints of consumers can easily be tracked in a unique way that the person will not even become aware of it. Virtually every single information about the person stemming from the person's interest, preference and some private information can easily be accessible to other third parties for them to use in their own way. Studies done by the Federal Trade Commission (FTC) discovered that 90 percent of web sites were gathering at least one type of identifiable information about their users such as name, e-mail address, location, while 57 percent were obtaining at least one type of demographic information, for example, gender and preferences with some websites going to the extent of gathering very sensitive details such as social security number and credit card details [5].

The sale of information has become very prevalent with the introduction of the dark web and here is a summary of how much some information gathered costs online from a TrendMicro research [6]:

- Very sensitive data such as banking credentials will cost around $1000. Health record and social media accounts sells for around $50 per record.

- Some basic information such as name, age, location, typically the kind of data that you need when registering for a site goes for about $0.50.

Even though this might not seem much, it must be noted that a number of these information floating around the dark web make it a very profitable venture for hackers, organizations and other people to use it to their advantage. It is clear that the financial reward of these kinds of information testifies to the reason why online gathering of consumer's data has taken an interesting turn for a lot of online services. And it is as a result of this that consumers now have a fear of how their private information is being used online.

With the release of Edward Snowden papers by WikiLeaks, it has been observed that the number of people using Virtual Private Networks (VPN) and browsers that ensure users anonymity online such as The Onion Router (TOR) and ZOHO have increased dramatically because as released by the WikiLeaks papers, these are the browsers that the NSA had major problems in decrypting [7].

When ex-CIA programmer Snowden released the classified files in 2013, in August alone the number of people using TOR skyrocketed from 500,000 users to 1,200,000 users [8]. Further studies which also show the ever-increasing number of users world-wide of TOR has been presented in [9]. Events like these have led to many people who initially were not interested in online privacy concerns to now question security practices of governments' branches that deal with the practice of looking into private lives of people by the use of the internet. Some governments are willing to prioritize the privacy of their citizens over the economic wellbeing of their citizens. For example, it was revealed recently that Verizon, a mobile telecommunication network gave US agencies backdoor access to be able to invade on the privacy of their users. The US agencies used this opportunity to spy on Angela Merkel, who is the German chancellor. Merkel cancelled Germany's partnership with Verizon which resulted in economic loss [10] when this vulnerability was uncovered. This proves how far governments are willing to go in order to protect the privacy of its people.

As a result of these occurrence, there have been a massive rise in technological start-ups companies that specialize in ensuring the anonymity or preventing the infringement of privacy of users. Experts encourage start-ups in this era of information technology to focus on solving problems on privacy. With an increasing number of devices and complex interconnectivity between them, attention is being shifted from how the data will be created and stored to how secure the data is. Consumers need to be able to dictate the use of their private data. The introduction of internet of things (IOT) which is defined as the inter-networking of physical devices, vehicles, buildings, and embedded systems such as electronics, sensors, actuators, and network connectivity which enables these objects to collect and exchange data has made the discussion on online privacy more imminent.

The Ghanaian telecom sector has seen an increasing number of mobile phones and other handset that can easily access the internet. This is as a result of the huge reduction in the cost of internet services. Internet has now become part of the everyday life of every Ghanaian. Ghana has a literacy rate of 71.5%. There are about 21 million educated people in Ghana and a survey conducted by researchers in [11] suggests that 51% of the people in Ghana use mobile phones. This is to say that getting access to internet is not too expensive for the average Ghanaian.

Research conducted from the user statistics of TOR, summarized in Fig. 1, shows the increase in the average Ghanaian internet consumer privacy since 2013 [12]. The later part of 2016 saw a spike in TOR usage. This could mainly be attributed to the cautious nature of the citizens during the presidential election period. The result in Fig. 1 supports the hypothesis that a concern has grown amongst the Ghanaian public about their online privacy and even though some experts might argue that citizens in third world countries like Ghana may not be concerned about online privacy, this data clearly tend to show otherwise that irrespective of the developmental stage of a country, privacy infringement is still a problem and hence Ghana is no exception to the rule.

The issue of online privacy in Ghana has become very important because businesses are now targeting Ghanaian consumers for e-Commerce transactions in their advertisements of business products and services. In most cases, consumers provide personal information during online transactions when registering on transactional websites such as Tonaton, OLX, Amazon, eBay, Alibaba and social media platforms like Facebook, Twitter, Instagram.

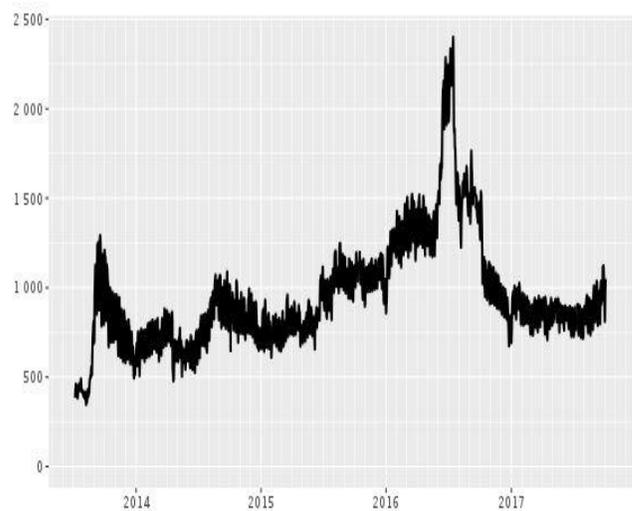

Fig. 1.   A graph showing the number of TOR browser users in Ghana [12].

Perkins and Annan [13] in their research work on evaluating factors that affects the adoption of online banking in Ghana found out that cookies in browsers provided personal information of customer when tracking customer's online surfing activities. They could not however evaluate the extent to which this breach of privacy affected online banking adoption by clients. This research seeks to investigate the extent to which the privacy concerns of online users in Ghana affect their willingness to share information online. The research also seeks to study if there is any relationship between the privacy concern and their willingness to transact business online.

### A. Principles of Privacy Development

The development of online privacy policies must address the perceived vulnerability of the online user and the perceived ability of the user to control his personal information online. The perceived ability to control is creating a condition for the consumer to believe that they can withhold personal information from being released online and also control what is being shared of them. This allows consumers to exercise their right to privacy. The perceived ability to control private information flow with consumers reduces their sense of vulnerability and their online privacy concerns [14].

The Federal Trade Commission addresses these issues with five core principles to promote privacy policy development by online content vendors [15]. These core principles are:

- Notice: Consumers ought to be made mindful of a Website's capacity to gather data and utilization practices before data is gathered.

- Choice: Consumers ought to be allowed to pick whether to partake or decline the collection of their own data.

- Access: Consumers ought to approach their own data and ought to have the capacity to rectify incorrect data.

- Security: There ought to be approaches to guarantee the respectability of information and as such, the anticipation of abuse of data ought to be set up.

- Redress: Enforcement ought to be set up to guarantee the above standards are maintained. Illustrations are self-controls (by the buyers) and government directions.

These guidelines allow online consumers to control the usage of their personal data. If these rules are met, the consumers will then feel safe and convinced to trust such websites to collect their personal information. Another way for consumers to trust websites to collect their information is the use of third-party seals of approval programs. Examples of such third parties are TRUSTe and Verisign. These programs confirm the website's respect for personal privacy [16].

## II. METHODOLOGY

This section investigates how the average Ghanaian online consumer uses the internet and how privacy issue affects the customers' willingness to use these online services again. The research further conducts a study on understanding how the new generation of internet users in Ghana perceives internet privacy and how it affects their internet usage.

The preferred method of information gathering was the questionnaire. Questions posed in the questionnaire were not compulsory, that is, respondents were free to ignore the question if they did not feel like answering. This idea was implemented as a way of making the answers as true as possible. No one was forced to answer any question. Respondents were also free to edit their answers. This was to allow answers that were selected in haste be reconsidered and changed at a later time. The disadvantage of this approach is that, it could lead to a lot of outliers and this could complicate data analysis.

About 120 individuals were invited of which 104 responded. The sample space comprised of Students and the working class adults who were avid internet users. This group was chosen because the young generation are the people who use the internet the most according to Perkins and Annan [13]. This category of users were also selected because of the need to get responses from people who were regular internet users, and that would mean they understood what exactly the research was about.

The main subject of the study was to evaluate how consumers are impacted by their perceived vulnerability to unauthorized gathering and use of personal information, and their perceived ability to control the manner in which their personal information is collected and used online. In view of this, the questions focused on the consumers' frequency of sharing information, what kind of information they shared and how safe they think their information is. Their response served as the basis to evaluate how Ghanaian consumers' perceived vulnerability affected their willingness to use the internet for e-Commerce transactions.

## III. RESULT AND DISCUSSIONS

As consumers' concerns about their online privacy grows rapidly, most activities on the internet could be eventually affected, thus, from e-Commerce to providing personal data and even general browsing/surfing on the internet. The first part of our study sought to find out how respondents share and look for information online. The results from the question were very conclusive that all of the people in the sampled class were internet users on various social platforms. They shared and received various types of information on the internet. Respondents indicated that, as seen in Fig. 1(a), they used the internet as their primary means of looking for information. 84.5% of respondent admitted that, they shared and researched for information online. When asked what kind of information they shared as summarized in Fig. 2(a), surprising, 99 responded to sharing information which contradicted the total number of 88 people who agreed to sharing information online. When the respondents were asked which kind of information they liked to share, 40 out of the 99 respondents indicated they liked to share personal information. Respondents were allowed to choose more than one option when submitting their response to this particular question. The major thing respondents seemed to like sharing was news with 69 out of 99 respondents indicating this preference. From Fig. 2(b), it is easy to notice that most of the things shared online by the sample group were non-personal details.

This was expected since the sample group was mostly young people who usually use the internet for reading news and browsing multimedia applications. The questions were further focused on details that had to do with their online vulnerability, that is, information shared that contained personal information. The responses are summarized in Fig. 3 and can be seen that 71.8% indicated they provided their personal details online. However, 72% of those who responded Yes to providing personal details online said that, they only provided it because that was the only means through which they could get access to the e-Commerce platforms they wanted to use.

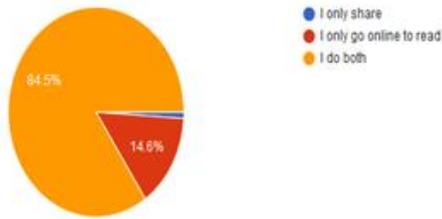

(a)

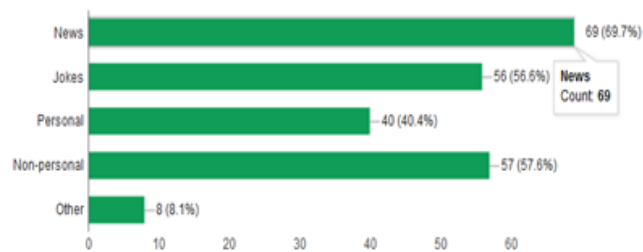

(b)

Fig. 2.   (a) Information sharing; (b) Information sharing by the sampled group.

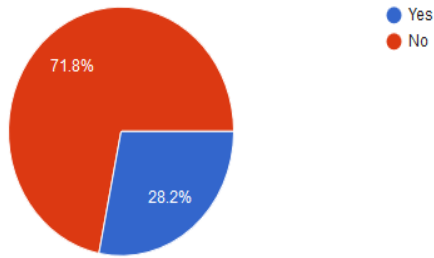

Fig. 3.   Personal details.

When the respondents were asked about their perceived vulnerability, 93.2% indicated that, the fear of the information they share been compromised online make them hesitant in sharing information. From the results summarized in Fig. 4 and 5, it is clear that most of the respondents were weary of the internet. They were afraid of what their information would be used for. This information differs from the assumptions some experts in literature make that internet users in third world countries don't care about their online privacy. It was assumed that the average Ghanaian did not care what kind of information they shared online, but from the results, it can be concluded that the new generation of Ghanaian internet users are aware of the dangers of the internet as supported by over 65% of respondents indicating their uncertainty of the safety of the internet (refer Fig. 5(a)) and as such their willingness to limit their information sharing (refer Fig. 5(b)).

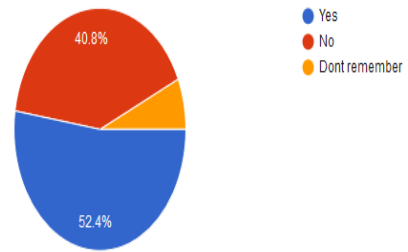

Fig. 4.   Perceived vulnerability.

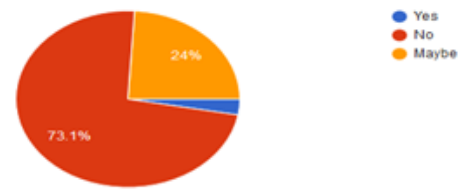

(a)

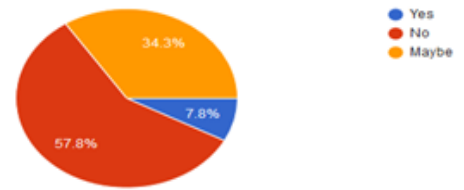

(b)

Fig. 5.   (a) Information sharing perspectives; (b) Perceived vulnerability concerns.

The respondents were further questioned on their knowledge of digital marketing and e-Commerce. Over 50% of them had heard of the phrase "Digital Marketing" but did not know exactly what it was whiles 70% had heard of e-Commerce and showed brilliant understanding of its concept.

After learning of their views on perceived vulnerability and their understanding of e-Commerce, it was decided to study how their perceived vulnerability affected their use of the internet. This part of the research questions sought to mainly inquire how the perceived vulnerability affected their interactions with e-Commerce websites and platforms that required them to share personal information in order to use their services. From the response gathered, 59.8% of respondents indicated they had made an online payment or transaction before as shown in Fig. 6 but only 10.1% of respondents felt that the internet was safe for making purchase orders and payment transactions as shown in Fig. 7(a).

Respondents were however, willing to shop online if e-Commerce was the only means of getting their products.

Have you ever made a purchase online?

102 responses

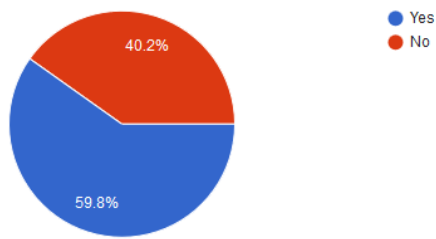

Fig. 6. Online purchasing responses.

Do you think online purchases are safe?

99 responses

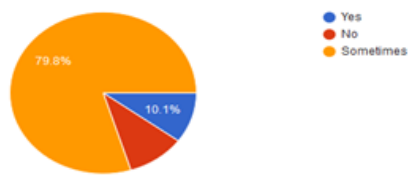

(a)

Would you shop online if it was the only way to get a product?

99 responses

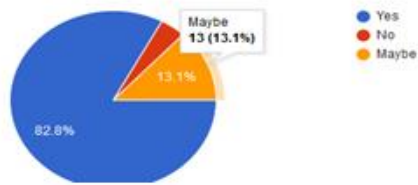

(b)

Fig. 7. (a) Views on the safety of e-Commerce; (b) Online purchasing responses.

In Fig. 7(a) and (b), there is a noticeable contrast in responses when it comes to the safety of e-commerce systems. In a bid to simplify the results, all 'Maybe' and 'Sometimes' answers were considered as negative answers. From this we can see that when asked about the safety of e-commerce platforms, most people did not trust them but when it was the only way to get a product most people said they would use them since they had no option.

Some of the reasons respondents gave for not trusting online e-Commerce sites were:

- Because such sites do not general feel safe for them to provide them with sensitive personal information.

- The respondents couldn't entirely trust online shops since they are unable to determine who is behind them.

- Some respondents also felt it was risky and that information provided could be used for something else. They therefore would prefer to deal directly with a supplier.

The results from this study show that Ghanaian users of e-Commerce platforms are highly concerned with their online privacy. About 71% of respondents are less willing to disclose personal information online. Despite their high privacy concerns and unwillingness to provide their personal information to websites, online privacy concerns did not negatively affect Ghanaians consumers' willingness to engage in e-Commerce transactions if the only means of getting the products was through an online transaction. Analysis of our questionnaire further indicated that this privacy concern of respondents could lead to three categories of internet users in Ghana, namely,

- *Users who are ready to provide their personal data*: In this case, users will not be required to enter or provide their personal details that the website requests from them. This may result in some surfing only through websites which do not request personal information or even in the cases where personal information is required, these users may provide false personal information in a bid to remain anonymous while using the website.

- *User who decline to adopt e-Commerce platform as a means of transacting business in Ghana*: Where their personal data will be required most of the time leading to an increase in the perceived vulnerability of internet users in Ghana. The possible increasing privacy concerns which could arise may affect negatively on e-Commerce transactions.

- *Users who refuse to use the internet and e-Commerce platforms*: This could be the worst case scenario where users would be unwillingness to use the internet and e-commerce platforms. This could happen when internet users have the feeling that even when they don't provide information, some attributes of theirs could still be tracked online. There are some users who could be extremely concerned about online privacy and would not even voluntarily use the internet.

People have tried to come out with frameworks that will reduce online privacy concern and through their analysis and study of the online privacy problem, a lot of strategies and measures have been suggested to solve or minimize the online privacy concerns including using an integrated whole of government approach [17] and framing a new internet regulatory regime [18].

This research would also propose some measures in addition to what have been proposed by Perkins and Annan [13] to address the online privacy concerns of the Ghanaian internet user.

The privacy concern can be reduced by increasing the users' perceived ability to control the collection and use of his sensitive personal information. This can be positively supported by the users' level of internet literacy and his level of awareness of social issues involving internet usage. The perceived control of user can be increased when businesses and legitimate marketers in Ghana who collect consumers' personal information obey the Privacy Law Act of Ghana which has been clearly stipulated in the constitution. This can be enhanced by online vendors by posting a comprehensive or detailed privacy policy conspicuously on the home page of their Website. This measure will increase the social awareness and alertness of internet users in Ghana. This could help in increasing the consumers' perceived ability to control information.

The procedural fairness in the collecting and usage of personal data collected from users by these e-Commerce platforms must also be made known to the users. This would further help increase users' trust in using these e-Commerce and online platforms [19]. Consumers should also be allowed to decide whether to choose to participate or be excluded in the collection of personal information before being taken. Consumers should have access to their own personal information and should be able to correct erroneous information. Policies to ensure the integrity of information and prevention of misuse of them should be clearly spelt out in the laws of Ghana.

Another way of reducing Ghanaians consumers' online privacy concerns, which could be derived from the implication of the results in this study, is by reducing the perceived vulnerability to information misuse and its consequences in e-Commerce. This can be achieved when users' perceived ability to control information collection and usage is increased as discussed early on. Since factors such as, accidental disclosure of personal information, hacking into networks and systems, weak identification and authentication systems and spam that require personal details, could affect users' perceived vulnerability. Users must be educated on these factors to improve their internet literacy levels. Data obtained from respondents in this study show that, only 5% of respondents knew about cookies and firewalls. It is therefore very important for online marketers and businesses in Ghana to educate online users of their platforms on the options and measures available for protecting their personal information.

## IV. CONCLUSION

From the study, it can be seen that Ghanaian consumers are highly concerned with their online privacy and they are less willing to disclose personal information online. Most of the respondents in this study indicated their preparedness to stay away from websites that require them to submit personal details such as bank account information and location information online. Even though their high privacy concern and unwillingness to provide their personal information to websites has not negatively affected the willingness to engage in e-Commerce transactions for now, their concerns about their online privacy could grow rapidly and this could adversely affect how they will adopt these ever-increasing e-Commerce platforms in Ghana in the long term if these concerns are not addressed soon.

The implication of the results in this study supports the need for online marketers and businesses to carefully study and understand these factors that increase online privacy concerns in Ghana and address them appropriately and urgently. Future research could develop and test a comprehensive privacy framework for the Ghanaian internet space. This is very crucial and critical because if not properly considered, online transactions and e-Commerce can lose its value and its impact on the Ghanaian economy would be adverse.